\documentclass[journal]{IEEEtran}
\ifCLASSINFOpdf
   \usepackage[pdftex]{graphicx}
\else
   \usepackage[dvips]{graphicx}
\fi
\usepackage{amsmath}
\usepackage{caption}
\usepackage{amssymb}
\usepackage{bm}
\usepackage{upgreek}
\usepackage{tabularx}
\usepackage{xcolor}
\begin{document}
%
\title{SAR Tomography via Nonlinear Blind Scatterer Separation}
%
%
%

\author{Yuanyuan Wang,~\IEEEmembership{Member,~IEEE,}
	and~Xiao Xiang Zhu,~\IEEEmembership{Senior~Member,~IEEE}
\thanks{This research was jointly supported by the European Research Council (ERC) under the European Union's Horizon 2020 research and innovation programme (grant agreement No. [ERC-2016-StG-714087], Acronym: \textit{So2Sat}), by the Helmholtz Association
through the Framework of Helmholtz Artificial Intelligence Cooperation Unit (HAICU) - Local Unit ``Munich Unit @Aeronautics, Space and Transport (MASTr)'' and Helmholtz Excellent Professorship ``Data Science in Earth Observation - Big Data Fusion for Urban Research'' and by the German Federal Ministry of Education and Research (BMBF) in the framework of the international future AI lab "AI4EO -- Artificial Intelligence for Earth Observation: Reasoning, Uncertainties, Ethics and Beyond" (\textit{Corresponding Author: Xiao Xiang Zhu})}
\thanks{Y. Wang and X. Zhu are with the Signal Processing in Earth Observation, Technical University of Munich, Munich, Germany, as well as with the Remote Sensing Technology Institute, German Aerospace Center. (email: yuanyuan.wang@dlr.de; xiaoxiang.zhu@dlr.de)}
}

%
%

\markboth{Journal of \LaTeX\ Class Files,~Vol.~14, No.~8, August~2015}%
{Shell \MakeLowercase{\textit{et al.}}: Bare Demo of IEEEtran.cls for IEEE Journals}
%



\maketitle

\begin{abstract}
\textcolor{blue}{\textbf{\textit{This work has been accepted by IEEE TGRS for publication.}}} Layover separation has been fundamental to many synthetic aperture radar applications, such as building reconstruction and biomass estimation. Retrieving the scattering profile along the mixed dimension (elevation) is typically solved by inversion of the SAR imaging model, a process known as SAR tomography. This paper proposes a nonlinear blind scatterer separation method to retrieve the phase centers of the layovered scatterers, avoiding the computationally expensive tomographic inversion. We demonstrate that conventional linear separation methods, e.g., principle component analysis (PCA), can only partially separate the scatterers under good conditions. These methods produce systematic phase bias in the retrieved scatterers due to the nonorthogonality of the scatterers’ steering vectors, especially when the intensities of the sources are similar or the number of images is low. The proposed method artificially increases the dimensionality of the data using kernel PCA, hence mitigating the aforementioned limitations. In the processing, the proposed method sequentially deflates the covariance matrix using the estimate of the brightest scatterer from kernel PCA. Simulations demonstrate the superior performance of the proposed method over conventional PCA-based methods in various respects. Experiments using TerraSAR-X data show an improvement in height reconstruction accuracy by a factor of one to three, {depending on the used number of looks}.
\end{abstract}


%
\IEEEpeerreviewmaketitle

\section{Introduction}
\begin{figure*}[h]
	\centering
	\includegraphics[width=0.37\textwidth]{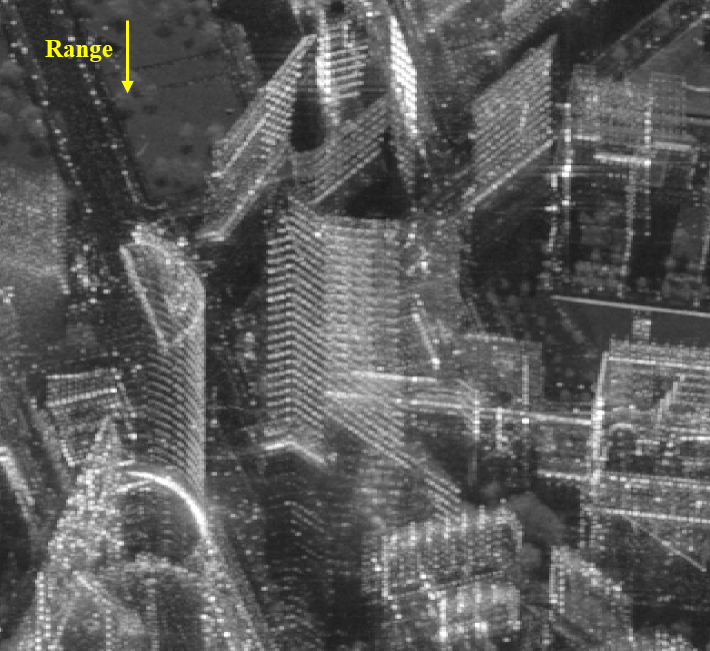} \includegraphics[width=0.45\textwidth]{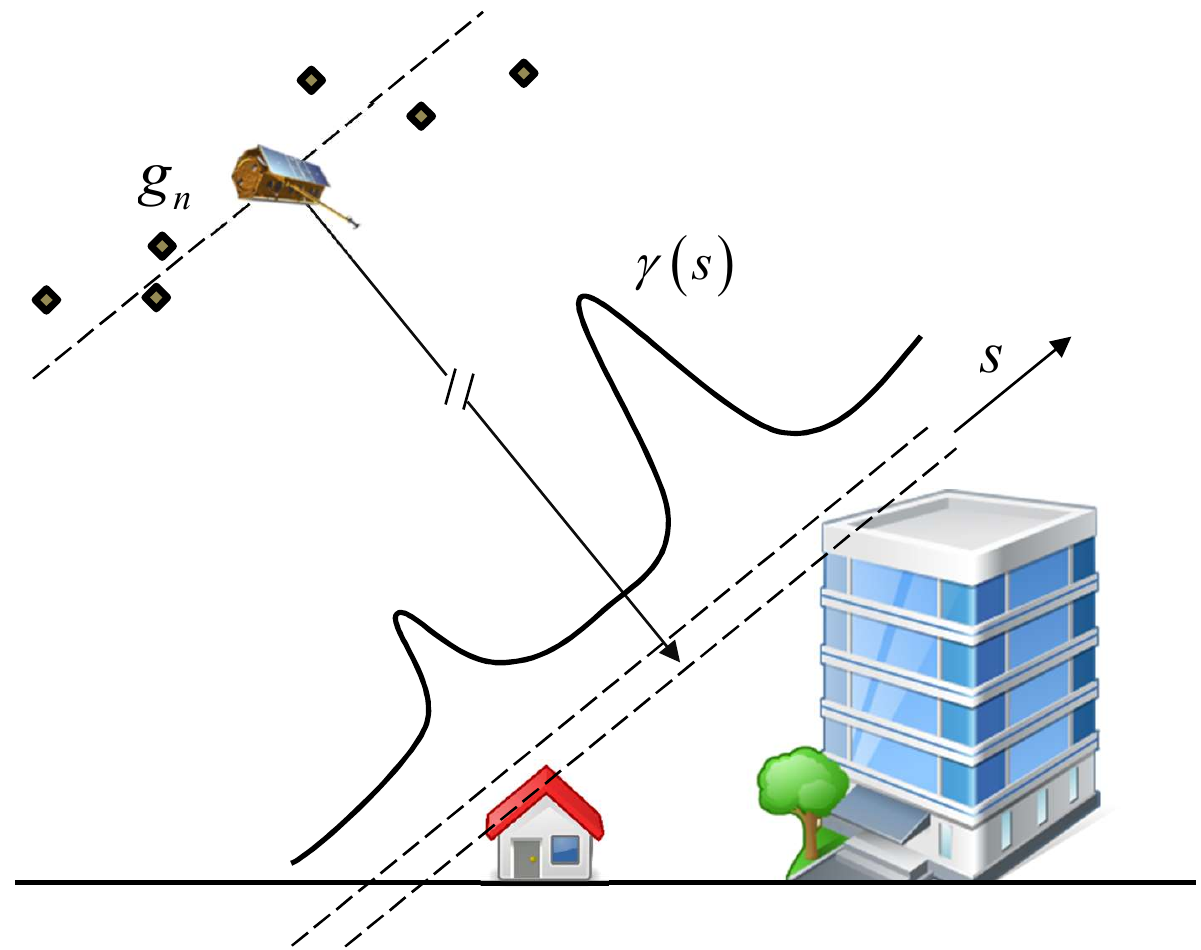}
	\caption{Left: a SAR intensity image showing the severe layover effect in a dense urban area (Berlin, Germany). The building ``collapsed'' towards the range direction. Right: the SAR imaging model at a fixed azimuth position. TomoSAR retrieves the reflectivity profile $\gamma \left(s\right)$ by building a synthetic aperture along the direction s using multiple acquisitions (indicated by the black diamonds) at different baselines.}
	\label{fig:imaging_model}
\end{figure*}

%
%
%
%
\IEEEPARstart{S}{ynthetic} aperture radar (SAR) interferometry is by far the most popular method for obtaining global digital elevation models, as well as for assessing long-term millimeter-level deformation over large areas of Earth’s surface. However, its side-looking imaging geometry causes inevitable layover in SAR images, especially in mountainous and urban areas. As an example, Fig. \ref{fig:imaging_model} (left) shows the layover phenomenon in a dense urban area. The buildings are layovered with the ground in front of them, appearing as if collapsed towards the sensor. As a result, the backscattering from the building facade and the ground is integrated into a single pixel during SAR image formation. Separating the contributions from different scatterers within one resolution cell has been the fundamental to many applications, such as urban 3-D reconstruction, and deformation monitoring in mountainous areas. It is typically solved by explicit inversion of the SAR imaging model in order to retrieve the scattering profile along the mixed dimension – that is \textit{elevation}. This process is known as SAR tomography (TomoSAR) or differential TomoSAR (D-TomoSAR), when the temporal motion of individual scatterers is also considered.

To further describe the layover effect, Fig. \ref{fig:imaging_model} (right) illustrates the SAR imaging model at a fixed azimuth position, where $\gamma \left(s\right)$ is the reflectivity profile along the elevation \textit{s}. TomoSAR builds a synthetic aperture along \textit{s} by utilizing multiple observations at different baselines. The multiple observations are usually acquired in a repeat-pass manner for spaceborne SAR sensors.
The continuous form of the SAR imaging model can be expressed as a Fourier transform at discrete frequencies, as follows.
\begin{equation} \label{ZEqnNum926374} 
	g_{n} =\int _{s}\gamma \left(s\right){\rm exp}\left(-j\frac{4\pi B_{n} }{\lambda r} s\right){\rm d}s ,  
\end{equation} 
where $g_{n} $ is the complex-valued observation acquired at baseline $B_{n} $, $\lambda $ and \textit{r }are the radar wavelength and range to the object, respectively. Equation \eqref{ZEqnNum926374} can be discretized as 
\begin{equation} \label{ZEqnNum514195} 
	\mathbf{g}=\mathbf{R}\bm{\upgamma} +\bm{\upvarepsilon},  
\end{equation} 
where $\mathbf{g}\in {\mathbb{C}}^{N} $ is the observation vector with \textit{N} elements, $\mathbf{R}\in {\mathbb{C}}^{N\times L} $ is the so-called \textit{steering matrix}, a discrete Fourier transform depending on the baselines and the \textit{L} discrete elevation values, $\bm\upgamma \in {\mathbb C}^{L} $ is the discretized reflectivity profile, and $\bm\upvarepsilon \in {\mathbb C}^{N} $ is the noise vector usually assumed to be independent and identically distributed (i.i.d.) complex circular Gaussian random variables. Due to its layover separation capability, TomoSAR has become the most competent multibaseline SAR interferometry (InSAR) techniques for various 3D reconstruction tasks. The technique has undergone extensive development since the era of very high-resolution spaceborne SAR sensors, e.g., TerraSAR-X. Some examples include improving the estimator by introducing regularization \cite{Fornaro2009Four-Dimensional,Zhu2010Very} using singular value decomposition, including nonlinear deformation parameters \cite{Zhu2011Let's,Reale2013Extension,Siddique2016Single-Look}, improving the detection of scatterers \cite{Pauciullo2012Detection,Ma2016Robust}, employing sparse reconstruction techniques to achieve super-resolution in the elevation reconstruction \cite{Zhu2010Tomographic, Zhu2014Superresolving}, as well as fusing SAR imaging geodesy \cite{Eineder2011Imaging} and TomoSAR inversion to obtain absolute geodetic TomoSAR \cite{Zhu2015Geodetic} point clouds. Comprehensive reviews of TomoSAR algorithms can be found in \cite{Zhu2014Superresolving,Zhu2011Very,Fornaro2014Tomographic}.

{High precision SAR tomographic reconstruction, and those requiring superresolution is computationally expensive than those traditional methods like persistent scatterer interferometry (PSI). } However, future SAR data will eventually converge to high resolution and wide coverage. For example, the German X-band high resolution wide swath SAR sensor is planned to launch in 2022 \cite{Bartusch2016HRWS}. At 0.25m resolution, it can cover 30$\mathrm{\times}$30 km${}^{2}$, which is tens of times larger than the coverage of the current staring spotlight mode of TerraSAR-X. Therefore, there is an emerging demand for developing computationally economical TomoSAR algorithms. This paper seeks to answer this question by summarizing the state of the art, enumerating the challenges, and proposing a new algorithm as well as directions for future development.

\subsection{Related work}
The higher computational cost of D-TomoSAR relative to PSI is due to its consideration of multiple scatterers instead of a single one, as in PSI. In D-TomoSAR, the parameters of all scatterers, which are the elevations and deformation parameters, must be searched simultaneously, exponentially increasing the solution space. In a rough approximation, the computational cost of D-TomoSAR is in the order of \textit{O}(\textit{L${}^{K}$}), assuming \textit{O}(\textit{L}) is the computational cost of a single-scatterer periodogram in PSI, \textit{L} is the discretization level of the parameter space, and \textit{K} is the number of scatterers. 

Reducing the computational cost of multi-dimensional optimization is often addressed by decomposing the optimization into several sub-problems whose optimization can be performed independently. This has been reflected in TomoSAR 
methods employing principle component analysis (PCA), such as the CAESAR algorithm \cite{Fornaro2015CAESAR:}. By separating the contributions from different scatterers, the computational complexity can be theoretically reduced to \textit{O}(\textit{KL}).

In a more general context, this type of decomposition problem is regarded as \textit{blind source separation }(BSS), which separates the contribution of individual sources without knowing the mixing matrix. In simple linear mixing case $\mathbf{x}=\mathbf{As}+\mathbf{n}$, the goal of BSS is to retrieve both the mixing matrix $\mathbf{A}$ and the source $\mathbf{s}$, given only the observations $\mathbf{x}$ subject to unknown measurement noise $\mathbf{n}$. As an alternative to model-based approaches, BSS methods have emerged as data-driven approaches in data science, where the unknown dynamics of the data are often hard to characterize. The most well-known BSS algorithms are inarguably those exploiting second order statistics of the data using PCA \cite{Belouchrani1997blind, Cardoso1998Blind}, and those exploiting higher order statistics using independent component analysis (ICA) \cite{Cardoso1990Eigen-structure,Cardoso1993Blind}. Extension has been created using kernel PCA (KPCA) \cite{Schoelkopf1997Kernel} to tackle nonlinear mixing models \cite{Martinez2003Nonlinear, Harmeling2003Kernel-based, Honeine2011Preimage}. A great deal of the attention has also been devoted to the joint diagonalization of a set of matrices \cite{Cardoso1993Blind,Cardoso1999High-order,Yeredor2002Non-orthogonal}, because the underlying key features of the mixed sources, e.g., statistical independence, can be expressed in terms of diagonal matrices \cite{Adali2014Source}. For example, PCA is essentially a diagonalization of the data covariance matrix. The list of literature on BSS is extensive, since the application of BSS covers vast research fields, including hyperspectral imaging, medical imaging, radar, electroencephalogram, audio processing, chemiometrics, and so on. 

Only a handful of studies has addressed BSS in InSAR. Most of them deal with the problem of polarimetric target decomposition using PCA and ICA \cite{Cloude1996review,Besic2015Polarimetric}. In the context of multibaseline InSAR, \cite{Fornaro2015CAESAR:} proposes using PCA to decompose layovered scatterers. However, \cite{DeZan2008Optimizing} argues that it is difficult to assign a physical interpretation to the eigenvectors of a data covariance matrix unless only a single dominant scatterer is present. Nevertheless, a reasonable result was demonstrated in \cite{Fornaro2015CAESAR:}. We discovered that PCA can only be applied in certain conditions, and proposed to employ KPCA in order to mitigate the errors \cite{wang_robust_2018}. 

\subsection{Contribution of this paper}
The BSS is different in SAR tomography than in conventional BSS applications, because the signal of our interest is often only the phase, instead of the whole complex-value. There has not been a systematic study of blindly separating complex multibaseline InSAR signals. The contribution of this paper is as follows. 

\begin{enumerate}
	\item  It provides a comprehensive review of the state of the art and systematically demonstrates the limitations of conventional methods in scatterers separation, mainly the low orthogonality between the scatterers. 
	
	\item  It proposes a nonlinear method to mitigate the phase bias caused by the limitations of the state-of-the-art algorithms. The proposed method employs KPCA to iteratively retrieve the direction of the currently most dominant scatterer. The contribution of the dominant scatterer in each iteration is deflated from the covariance matrix after estimating the intensity using the Rayleigh quotient. The algorithm is fully nonparametric. 
	
	\item We also designed a robust workflow for real data processing.
\end{enumerate}

\subsection{Notations}

This paper make use of a bold capital letter to denote a matrix, bold lowercase letter for a vector, and italic letters (both upper and lowercase) for a scalar. Frequently appearing quantities and notations in the paper are listed in Table \ref{tab:notation}.


\begin{table}[h]
	
	\caption{Symbols and notations}
	\begin{tabular}{p{1.8cm} p{6.0cm}} \hline 
		$N\in {\mathbb N}$ & Number of observations (images in our case) \\  
		$M\in {\mathbb N}$ & Number of samples \\ 
		$L\in {\mathbb N}$ & Discretization level in the elevation direction \\  
		$K\in {\mathbb N}$ & Number of scatterers \\ 
		$\mathbf g\in {\mathbb C}^{N} $ & Multibaseline InSAR observations \\
		$\bm\gamma\in {\mathbb C}^{L} $ & Reflectivity profile along elevation \\
		$\mathbf G\in {\mathbb C}^{N\times M} $ & \textit{M} samples of multibaseline InSAR observations \\ 
		$\mathbf R\in {\mathbb C}^{N\times K} $ & Forward model matrix (aka. Steering matrix), for \textit{K} scatterers  \\ 
		$\mathbf C\in {\mathbb C}^{N\times N} $ & Complex-valued covariance matrix \\  
		$\bm\Phi \in {\mathbb C}^{N\times N} $ & Diagonal matrix of the modeled phase \\  
		$\left\langle ,\right\rangle $  & Inner product \\ 
		$\det \left(\cdot \right)$ & Matrix determinant \\ 
		$\left|\cdot \right|$ & Absolute value (element-wise if on vector or matrix) \\ \hline 
	\end{tabular}
\label{tab:notation}
\end{table}%

\section{ Problem formulation}
\subsection{Mathematical model}
\noindent \textbf{Fully coherent model }

Assuming the reflectivity profile is coherent over the multiple acquisitions at different baselines or acquisition time, the discrete SAR imaging model can be expressed by equation \eqref{ZEqnNum514195}. In an urban area, the profile $\bm\upgamma$ usually consists of a few scatterers. For a \textit{K-}scatterer profile, the imaging model can be simplified to
\begin{equation} \label{3)} 
	\mathbf g=\left[\begin{array}{cccc} {\mathbf r_{1} } & {\mathbf r_{2} } & {\cdots } & {\mathbf r_{K} } \end{array}\right]\left[\begin{array}{c} {\gamma _{1} } \\ {\gamma _{2} } \\ {\vdots } \\ {\gamma _{K} } \end{array}\right]+\bm\upvarepsilon,  
\end{equation} 
where $\mathbf r_{k} $ is the column steering vector corresponding to the \textit{k}th scatterer, and $\gamma _{k} $ is the complex-valued amplitude of the \textit{k}th scatterer.

The covariance matrix of the observations ${\mathbf g}$ is as follows
\begin{align} \label{4)} 
	\mathbf C_{\mathbf{gg}} &=\mathrm E\left\{\mathbf{R}\bm\upgamma \bm\upgamma^{H} \mathbf{R}^{H} +\bm\upvarepsilon \bm\upvarepsilon ^{H} \right\} \nonumber \\ 
	&=\mathbf R \mathrm E\left\{\bm\upgamma \bm\upgamma ^{H} \right\}\mathbf R^{H} + \mathrm E\left\{\bm\upvarepsilon \bm\upvarepsilon ^{H} \right\}, 
\end{align} 
where $\left(\cdot \right)^{H} $ is the conjugate transpose operator. The profile $\bm\upgamma $ can be assumed to be uncorrelated, which leads $\mathrm E\left\{\bm\upgamma \bm\upgamma ^{H} \right\}$ to a diagonal matrix with the expected intensity of individual scatterers. The observation covariance matrix can be simplified to
\begin{equation} \label{ZEqnNum639696} 
	\mathbf C_{\mathbf{gg}} =\sum _{k=1}^{K}\sigma _{k}^{2} \mathbf r_{k} \mathbf r_{k}^{H}  +\sigma _{\varepsilon }^{2} \mathbf I,
\end{equation} 
where $\sigma _{k}^{2} $ is the expected intensity of the \textit{k}th scatterer, and $\sigma _{\varepsilon }^{2} \mathbf I$ accounts for the covariance matrix of the noise. Without losing generality, we can assume the steering vectors $\mathbf r_{k} $ are all normalized.

Such a scattering model can resemble the layover of very coherent DSs (or even PSs). However, the intensities of the scatterers across different spatial samples are not assumed to be deterministic. They are assumed to be Gaussian scatterers so that equation \eqref{ZEqnNum639696} holds. This is quite common in urban areas, where many adjacent scatterers on facades are very coherent over different (temporal and spatial) baselines, yet their intensities are stochastic among the adjacent samples. This model is known in radar jargon as the Swerling II model \cite{Richards2010Principles}.

\hfill

\noindent \textbf{Decorrelating DS model}

In a more general case, the decorrelation of the reflectivity profile $\bm\upgamma $ due to geometric or temporal baselines should be considered. For a \textit{K}-scatterers case, the imaging model can be formulated as follows:
\begin{equation} \label{ZEqnNum442069} 
	\mathbf C=\sum _{k=1}^{K}\bm\Phi _{k} \mathbf C_{k} \bm\Phi _{k}^{H}  +\sigma _{\varepsilon }^{2} \mathbf I,
\end{equation} 
where $\bm\Phi _{k}$ is a diagonal matrix constructed from $\mathbf r_{k}$, and $\mathbf C_{k} $ is the positive real-valued decorrelation covariance matrix of the \textit{k}th scatterer. Under this type of model, \textit{K} covariance matrices with a total of \textit{N}$\mathrm{\times}$\textit{N}$\mathrm{\times}$\textit{K }parameters are to be estimated, instead of the only \textit{K} intensities in the fully coherent model. The fully coherent model is a special case of the decorrelating model where $\mathbf C_{k} $ degenerates to all-ones matrices. Solving the full covariance matrices as well as the steering vectors in equation \eqref{ZEqnNum442069} using BSS remains at a challenge. We only found solutions under certain specific conditions, e.g., constant coherence. Therefore, the decorrelating DS model is not considered in this paper. 

The following content of this article will be based on the fully coherent model expressed in equation \eqref{ZEqnNum639696}. The goal of BSS is to retrieve the steering vectors $\mathbf r_{k} $ of the scatterers, as well as the intensities $\gamma _{k} $, from the observations \textbf{g} without performing TomoSAR, i.e., inverting the SAR imaging model and detecting the scatterer's location. Depending on the applications, permutation, scaling, or a constant common phase offset of the steering vectors may all lead to valid solutions. For example, all  three options are valid for 3D reconstruction, because only the relative position of the two scatterers matters in practice. In the following content, we define that the steering vectors are sorted in a descending order according to their corresponding intensity, and the steering vectors have a unit norm.

\subsection{PCA in TomoSAR scatterer separation}
\label{sec:limit_pca_pca}
The common idea of BSS algorithms is to exploit statistical independency of the sources. By assuming statistical independency, one can seek a diagonalization of the covariance matrix of the observation. PCA \cite{Jolliffe1986Principal} diagonalizes the data by converting the data into a set of orthogonal basis, known as principle components, whose variances are subsequently maximized. Performing PCA on a data covariance matrix that is semi-positive definite is equivalent to eigenvalue decomposition (EVD). For our BSS problem, we denote the EVD of a covariance matrix of the observations as follows.
\begin{equation} \label{7)} 
	\mathbf C_{\mathbf{gg}} =\mathbf{UDU}^{H} ,  
\end{equation} 
where $\mathbf{U}$ and $\mathbf{D}$ are the eigenvectors and the diagonal matrix, respectively. 

Although \cite{DeZan2008Optimizing} mentions that it is difficult to assign exact meaning to the eigenvectors, $\mathbf{U}$ can be approximated as the steering vectors of the individual scatterers under a \textit{strong orthogonality condition} among all the scatterers, as will be explained in Section \ref{sec:limit_pca_tomosar}. Such approximation was employed in the CAESAR algorithm \cite{Fornaro2015CAESAR:}. However, if the multibaseline SAR observation has a single dominant scatterer, the eigenvector corresponding to the largest eigenvalue of the data covariance matrix is an unbiased estimate of the steering vector of the scatterer. This has been mentioned in \cite{DeZan2008Optimizing,Even2016Study}. We show a short proof as follows.

\hfill

\noindent \textbf{Proof:}   For a single scatterer, its theoretical covariance matrix is of the form $\mathbf C_{\mathbf{gg}}=\bm\Phi _{1} \mathbf C_{1} \bm\Phi _{1}^{H} $ according to equation \eqref{ZEqnNum442069}, where $\mathbf C_{1} $ is a real-valued decorrelation matrix, and $\Phi _{1} $ is the diagonal matrix of the steering vector. 

\begin{enumerate}
	\item $\mathbf C_{\mathbf{gg}} $ can be expressed as $\mathbf C_{\mathbf{gg}}=\bm\Phi _{1} \mathbf C_{1} \bm\Phi _{1}^{H}  =\bm\Phi _{1} \left(\mathbf{VDV}^{H} \right)\bm\Phi _{1}^{H} $, where $\mathbf{V}$ and $\mathbf{D}$ are the eigenvectors and the eigenvalues matrix of $\mathbf C_{1}$, respectively. 
	
	\item Let $\mathbf U=\bm\Phi _{1} \mathbf V$. It can be shown that $\mathbf U=\bm\Phi _{1} \mathbf V$ is orthogonal, hence $\mathbf{UDU}^{H} $ is the EVD of $\mathbf C_{\mathbf{gg}}$.
	
	\item Since $\mathbf C_{1} $ is always real symmetric and positive definite, $\mathbf{V}$ and $\mathbf{D}$ are both real, according to \cite{Ayres1967Theory}.
	
	\item  Since $\mathbf{V}$ is real, the phase of each column of $\mathbf U$ is identical to the diagonal of $\bm\Phi_{1} $.
\end{enumerate}

\subsection{Limitations of PCA}
\label{sec:limit_pca_tomosar}
\noindent \textbf{Inseparability of nonorthogonal sources}

\begin{figure}[h]
	\centering
	\includegraphics[width=0.48\textwidth]{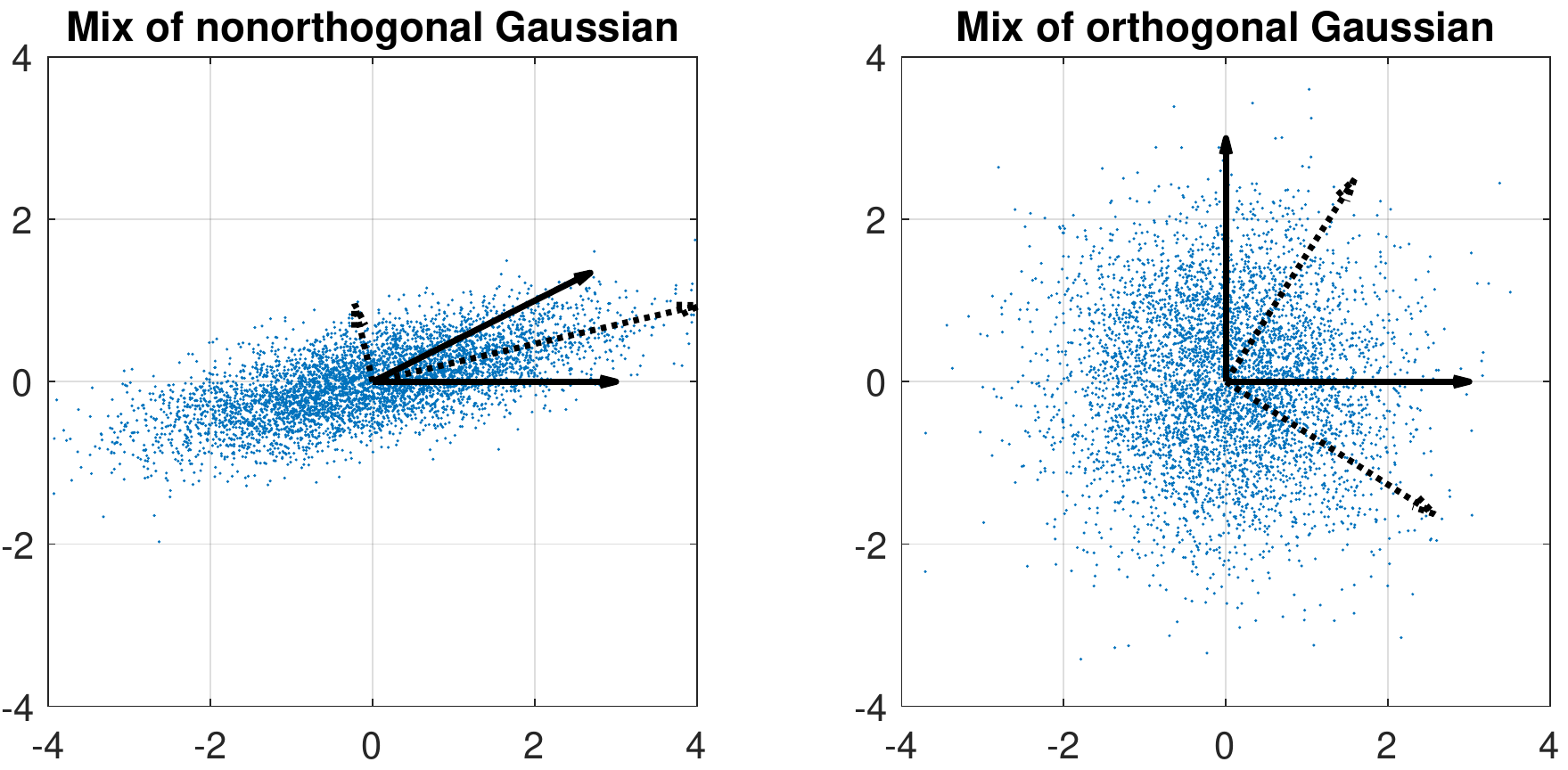}
	\caption{{Mixture of nonorthogonal (left) and orthogonal (right) 2-D Gaussians of identical standard deviation (set to 1). The solid arrows are the true directions of the sources, and the dashed arrows are the directions extracted by PCA. The length of the arrows is three times that of the (estimated) standard deviation. The left subfigure shows that a nonorthogonal mixture of Gaussian sources cannot be separated by linear PCA, while the right subfigure shows that an orthogonal mixture of Gaussian sources with the same variance can also not be unmixed by linear PCA.}}
	\label{fig:limit_pca_mag}
\end{figure}

The limitation of PCA is two-fold. On one hand, it assumes a linear combination of orthogonal basis. Hence, it is unable to fully recover nonorthogonal basis. On the other hand, the directions of the sources are indeterminable if the variance of the individual sources are identical. These can be exemplified by a 2-D Gaussian mixture shown in Fig. \ref{fig:limit_pca_mag}. The two subfigures are mixtures of nonorthogonal and orthogonal 2-D Gaussian sources of unit variance, respectively. The solid arrows in the figure are the true directions of the sources, and the dashed arrows are the directions estimated by PCA. The length of the arrows is three times that of the estimated or the true standard deviation. The left subfigure demonstrates that a nonorthogonal mixture of Gaussian sources cannot be separated by linear PCA. Both the direction and the variance of the sources were not correctly estimated. In contrast, the right subfigure shows that orthogonal mixture of Gaussian sources of identical variance can be separated by linear PCA, but subject to a random angle offset. 

Generalizing to our BSS problem in multibaseline InSAR, the performance of PCA is poor when the orthogonality of the two scatterers is low. We discovered the following three common causes of low orthogonality in multibaseline InSAR:
\begin{enumerate}
	\item a low number of images,
	\item a short distance between the two scatterers, i.e., either close to or shorter than the Rayleigh resolution, and 
	\item similar amplitude among the scatterers, which causes a severe interference.
\end{enumerate}
The degradation of performance is reflected in the phase bias (as well as in the amplitude bias, which is not of our primary interest) of the extracted steering vectors of individual scatterers. An example of this phase bias of the estimated steering vector is shown below.

\hfill

\noindent \textbf{Systematic phase bias}

\begin{figure}[h]
	\centering
	\includegraphics[width=0.49\textwidth]{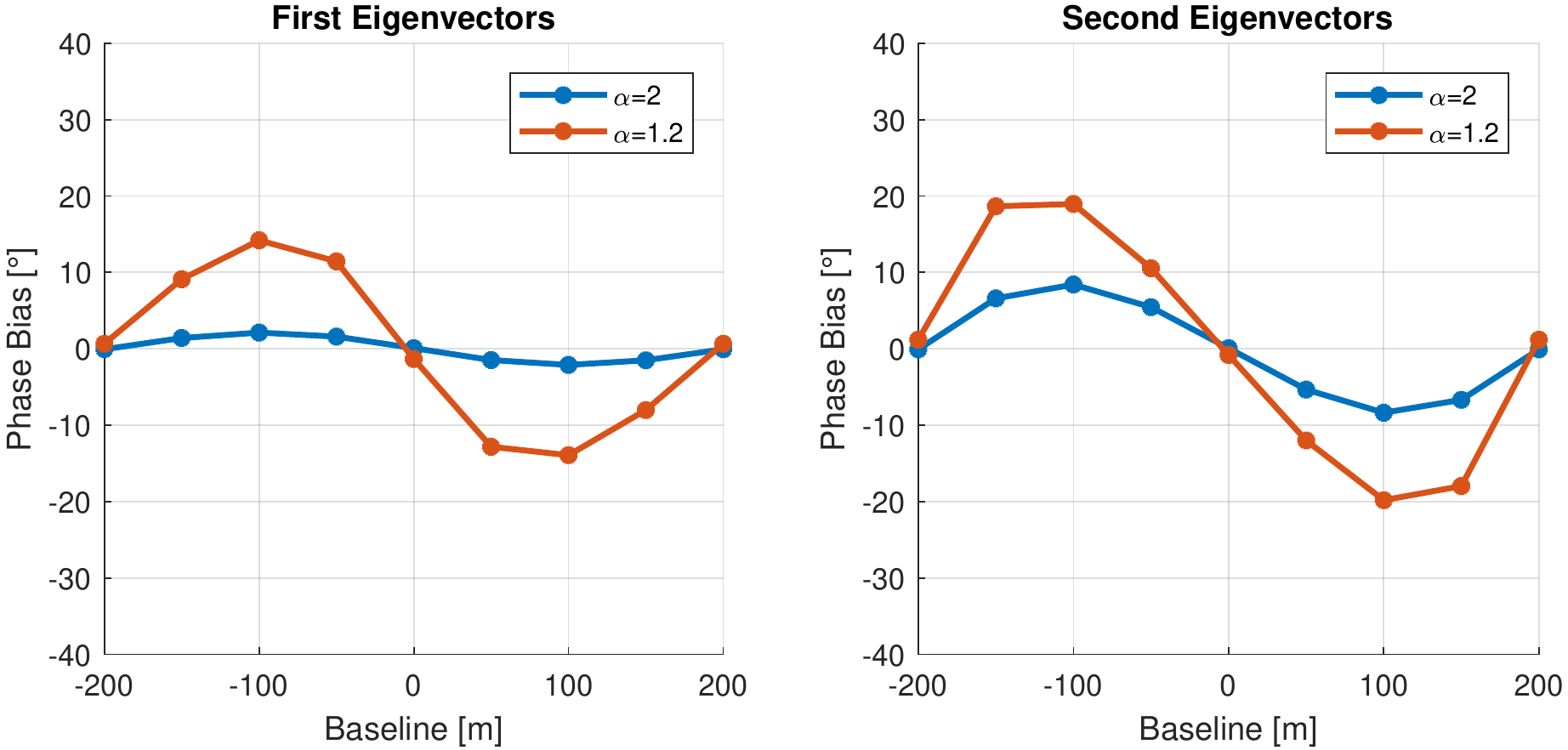}
	\caption{Phase bias of the first and second eigenvectors with respect to the perpendicular baseline, at different amplitude ratios ($\alpha=2, 1.2$) of the two scatterers. The larger amplitude ratio between the scatterers gives less phase bias, hence better separability of the scatterers. The bias of the second eigenvector is in general larger than the first one.}
	\label{fig:limit_pca_phase}
\end{figure}

Fig. \ref{fig:limit_pca_phase} demonstrates the phase bias of the eigenvectors extracted by PCA from a simulation of a two-scatterer mixture. The number of observations was set to nine. The perpendicular baselines are equally distributed from -200 m to 200 m, with other parameters similar to those of TerraSAR-X. This leads to a Rayleigh resolution of 27m in this simulation. The two scatterers are set to locate at 40 m and 80 m. No noise was included in the simulation. The two curves in each subplot in Fig. \ref{fig:limit_pca_phase} correspond to an amplitude ratio $\alpha$ of 1.2 or 2 between the two scatterers, respectively. The phase bias appears as a periodic undulation with respect to the perpendicular baseline. Consistent with the analysis shown in Fig. \ref{fig:limit_pca_mag} (right), the phase bias increases as the amplitude ratio of the two scatterers approaches one. The bias of the second eigenvector is larger than that of the first. In this example, the maximum phase bias of the eigenvectors for $\alpha=2$ is about 4$\mathrm{{}^\circ}$ and 15$\mathrm{{}^\circ}$. This corresponds to about 0.3m and 1m bias, respectively, in elevation. Such precision is sufficient for certain applications. However, as the amplitude ratio and SNR vary, large systematic phase bias can be expected from conventional PCA-based methods.

\section{Nonlinear Blind Scatterer Separation}
Based on the discussion in Section \ref{sec:limit_pca_tomosar}, the performance of PCA degrades as the orthogonality of the two scatterers reduces. 
As the amplitude ratio of the scatterers in the data cannot be altered, increasing the dimensionality of the data is the fundamental strategy to improve the orthogonality in our proposed algorithm. As it is usually infeasible to increase the number of images, the proposed method artificially increases the dimension of the data by projecting them into a higher dimensional space through nonlinear transformation. The BSS is then performed in the transformed higher dimensional space. The proposed algorithm employs the \textit{kernel trick} \cite{Hofmann2008Kernel} to perform the BSS in a Hilbert space without explicitly evaluating in the higher dimensional space. This can be easily realized using KPCA. 

\begin{figure}[h]
	\centering
	\includegraphics[width=0.35\textwidth]{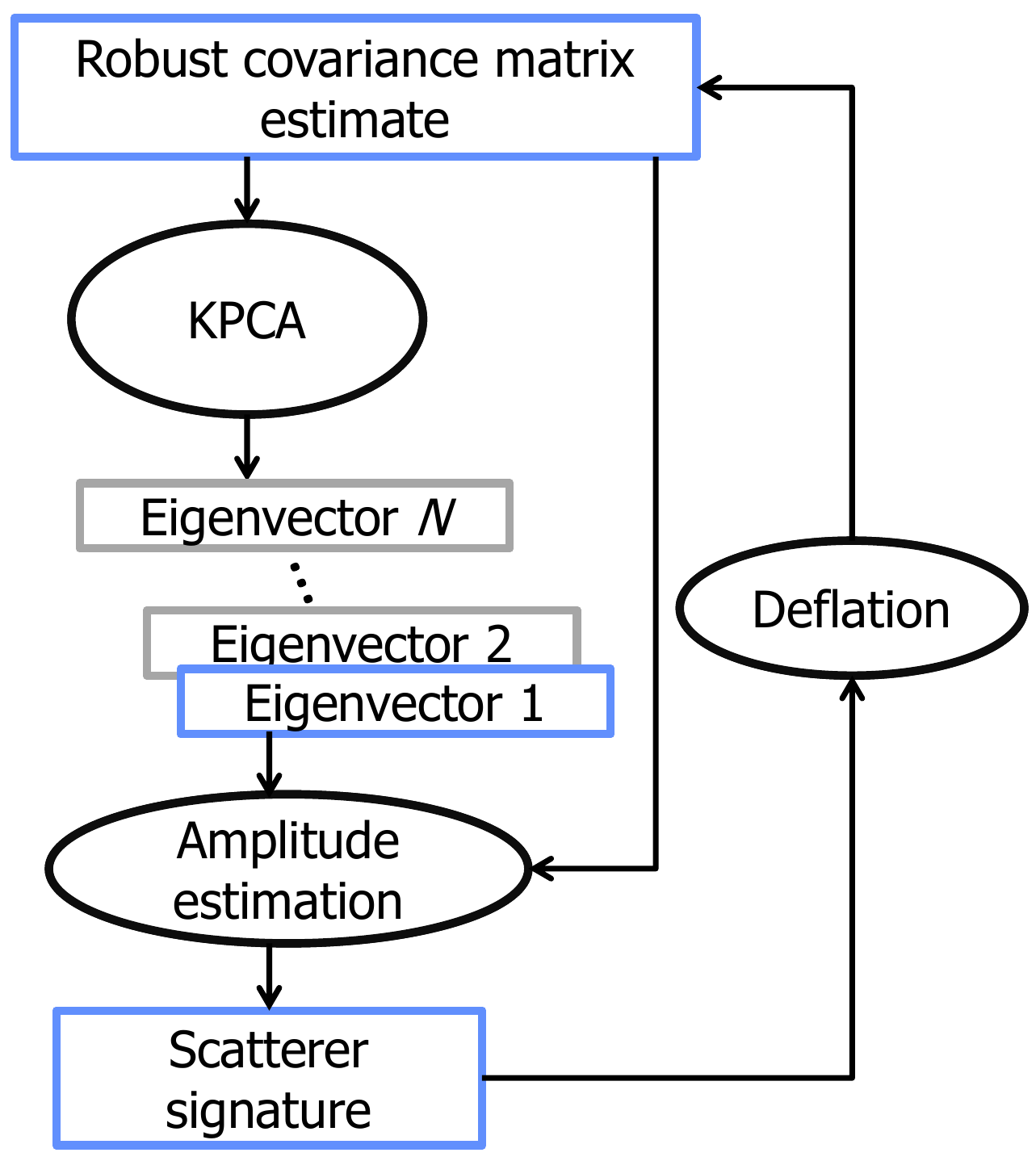}
	\caption{The flowchart of the proposed nonlinear blind scatterer separation algorithm.}
	\label{fig:workflow}
\end{figure}

The proposed algorithm performs in an iterative manner. At each iteration, it extracts and demodulates the dominant scattering contribution from the data covariance matrix, until no significant scattering is left or the predefined maximum number of scatterers is reached. The flowchart of the proposed algorithm is shown in Fig. \ref{fig:workflow}. The algorithm is explained in detail in the following sections.

\subsection{Dominant scatterer extracting via kernel PCA}

We will denote $\mathbf G\in {\mathbb C}^{N\times M} $ as the matrix of \textit{M} ergodic samples of the multibaseline observation $\mathbf{g}$. The scatterers of each sample are assumed to be stationary random variables, i.e., identical phase centers of the scatterers. In real data, we should imagine pixels with similar layover configurations, e.g., a row of pixels on the same floor of a fa\c{c}ade which layovers with flat ground. As the discussion of the sample selection is out of the scope of this paper, we assume the aforementioned condition is met in this article. This allows us to make use of the second order statistics, i.e., the covariance matrix, of the data. Mathematically, the singular vectors of $\mathbf G$ are identical to the eigenvectors of the sample covariance matrix $\hat{\mathbf C}=M^{-1}\mathbf{GG}^{H} $. In the following derivation, we will denote $\mathbf{C}$ as the input data matrix of BSS, instead of $\mathbf{G}$.

PCA performs a linear separation in the original data space, i.e., in $\mathbf{C}$. As described above, KPCA is a nonlinear generalization of PCA. The nonlinear separation is achieved through a linear separation on a nonlinearly transformation $\Psi$ of the data \cite{Schoelkopf1997Kernel,Harmeling2003Kernel-based}, which can be expressed as follows. 
\begin{equation} \label{8)} 
	\Psi :{\mathbb C}^{N} \to F,\; \mathbf c\to \Psi \left(\mathbf c\right),  
\end{equation} 
where \textit{F} is the transformed vector space which may have arbitrary dimension, and $\mathbf{c}$ denotes the columns of $\mathbf{C}$. Let us denote the transformed data as $\bm\Psi _{\mathbf c} =\left[\Psi \bm\left(\mathbf c_{1} \bm\right),\; \Psi \left(\mathbf c_{2} \right),\cdots ,\; \Psi \left(\mathbf c_{N} \right)\right]$. The KPCA of $\mathbf{C}$ basically finds the EVD of the covariance matrix in the transformed space, which is
\begin{equation} \label{9)} 
	\mathbf C_{\Psi \Psi } =\bm\Psi _{\mathbf c} \bm\Psi _{\mathbf c}^{H} =\mathbf U_{\Psi \Psi } \mathbf D_{\Psi \Psi } \mathbf U_{\Psi \Psi }^{H} .  
\end{equation} 
As the nonlinear transformation can have infinite dimension, EVD is never explicitly evaluated: they are indirectly evaluated through the kernel trick. It assumes that a Hilbert space of \textit{F} can be represented by a certain kernel function of the input data space, i.e.,
\begin{equation} \label{ZEqnNum324159} 
	\kappa \left(\mathbf c_{i} ,\mathbf c_{j} \right)=\Psi \left(\mathbf c_{i} \right)^{H} \Psi \left(\mathbf c_{j} \right),
\end{equation} 
where $\kappa \left(\cdot \right)$ is a kernel function, and $\mathbf c_{j} $ refers to the \textit{j}th\textit{ }sample (column) of $\mathbf{C}$. For convenience, we define the kernel matrix $\mathbf K\in {\mathbb C}^{N\times N} $ of the transformed data as 
\begin{equation} \label{11)} 
	\mathbf K=\bm\Psi _{\mathbf c}^{H} \bm\Psi _{\mathbf c}.  
\end{equation} 
Hence, each element of the kernel matrix can be easily found via equation \eqref{ZEqnNum324159}. The EVD of the kernel matrix is immediately available as follows.
\begin{equation} \label{ZEqnNum811987} 
	\mathbf{KV}=\mathbf{VS},  
\end{equation} 
where $\mathbf{V}$ is the eigenvectors, and $\mathbf{S}$ is the matrix of eigenvalues. Multiplying both sides of equation \eqref{ZEqnNum811987} by $\bm\Psi_{\mathbf c}$ gives
\begin{equation} \label{13)} 
	\bm\Psi_{\mathbf c} \bm\Psi _{\mathbf c}^{H} \left(\bm\Psi _{\mathbf c} \mathbf V\right)=\left(\bm\Psi _{\mathbf c} \mathbf V\right)\mathbf S,  
\end{equation} 
which implies that $\bm\Psi _{\mathbf c} \mathbf V$ and $\mathbf{S}$ are the eigenvectors and eigenvalues of the covariance matrix $\mathbf C_{\Psi \Psi } =\bm\Psi _{\mathbf c} \bm\Psi _{\mathbf c}^{H} $, respectively.

By properly choosing the kernel function, $\bm\Psi _{c} \mathbf V$ shall represent the space spanned by individual scatterers, or at least one of them. Hence, the data projected onto these eigenvectors shall represent the array manifold, i.e., the steering vectors. By taking and normalizing the first \textit{K }eigenvectors of $\mathbf C_{\Psi \Psi } $, the orthogonal projection basis in the higher dimension can be obtained as follows.
\begin{equation} \label{ZEqnNum371798} 
	\bm\Xi =\bm\Psi _{\mathbf c} \mathbf V_{1\sim K} \mathbf S_{1\sim K}^{-{1\mathord{\left/ {\vphantom {1 2}} \right. \kern-\nulldelimiterspace} 2} }, 
\end{equation} 
where $\mathbf V_{1\sim K} $ and $\mathbf S_{1\sim K}^{} $ denote the first $K$ columns of $\mathbf{V}$ and $\mathbf{S}$. Although, the eigenvectors and eigenvalues of $\mathbf C_{\Psi \Psi } $ appear in equation \eqref{ZEqnNum371798}, the EVD of $\mathbf C_{\Psi \Psi } $ is never explicitly evaluated. Only the calculation of the kernel matrix is required \cite{Harmeling2003Kernel-based,Wang2012Kernel} when projecting the data onto this basis. The projected data can be obtained as follows.
\begin{equation} \label{ZEqnNum873526} 
	\mathbf Y=\bm\Psi _{\mathbf c}^{H} \bm\Xi =\mathbf K \mathbf V_{1\sim K} \mathbf S_{1\sim K}^{-{1\mathord{\left/ {\vphantom {1 2}} \right. \kern-\nulldelimiterspace} 2} } . 
\end{equation} 
The phase of the first column of $\mathbf{Y}$ is extracted as the estimate of the steering vector of the first scatterer.

\subsection{ Selection of the kernel}
\label{sec:selection_kernel}
We seek a proper kernel function that allows the energy of different scatterers, or at least the most dominant scatterer, to be well localized in the individual columns of the transformed data $\mathbf{Y}$. In order to achieve that, the kernel should be able to transform the elliptically distributed scatterers into a linear coordinate where they can be separated. Kernels that have been extensively discussed in various applications are polynomial and Gaussian kernels, as they are both able to transform a radial basis to a near-linear basis.

\hfill

\noindent \textbf{Polynomial kernel}

A polynomial kernel is usually defined as follows:
\begin{equation} \label{16)} 
	\kappa \left(\mathbf c_{i}^{} ,\mathbf c_{j} \right)=\left(\mathbf c_{i}^{H} \mathbf c_{j} +1\right)^{d} ,  
\end{equation} 
where $\mathbf c_{i}^{H} \mathbf c_{j} $ emphasizes the \textit{angular difference} between the data, while the order \textit{$d\in {\mathbb R}^{+} $ }introduces nonlinearity and increases the dimension. For integer polynomial order \textit{d}, it computes a dot product in the space spanned by all monomials of degree \textit{d} in the input coordinates \cite{Harmeling2003Kernel-based,Hofmann2008Kernel}. For example, the second order polynomial kernel of a mixture of two sources \textit{s}${}_{1}$ and \textit{s}${}_{2}$ will be spanned by the monomials $\left\{s_{1}^{2} s_{2}^{2} ,\; s_{1}^{2} s_{2}^{} ,\; s_{1}^{} s_{2}^{2} ,\; s_{1}^{2} ,\; s_{2}^{2} ,\; s_{1}^{} s_{2}^{} ,\; s_{1}^{} ,\; s_{2}^{} \right\}$. The dimension under non-integer polynomial orders can be very high. It is obvious that polynomial orders greater than two are not feasible for our BSS problem, since they will introduce artificial scatterers with higher order phase term into the mixture. Such higher order scatterers have effective elevations of multiple times of the original ones, which may cause phase ambiguity. 

\hfill 

\noindent \textbf{Gaussian kernel}

A Gaussian kernel is defined in equation \eqref{ZEqnNum406817} below.  
\begin{equation} \label{ZEqnNum406817} 
	\kappa \left(\mathbf c_{i}^{} ,\mathbf c_{j} \right)=\exp \left(-{\left\| \mathbf c_{i} -\mathbf c_{j} \right\| _{2}^{2} \mathord{\left/ {\vphantom {\left\| \mathbf c_{i} -\mathbf c_{j} \right\| _{2}^{2}  \left(2\sigma ^{2} \right)}} \right. \kern-\nulldelimiterspace} \left(2\sigma ^{2} \right)} \right). 
\end{equation} 
Unlike the polynomial kernel, a Gaussian kernel emphasizes the Euclidean distance between the data. The standard deviation $\sigma$ of the Gaussian kernel depends on the \textit{Euclidean distance} between the scatterers. Ideally, it should be smaller than \textit{inter}-scatterer distances while larger than \textit{inner}-scatterer distances. Although, the scatterers are yet unknown, one can still estimate the inner scatterer distance by finding the minimum distance between samples \cite{Wang2012Kernel}. Following the same idea, we propose an estimator which is expressed in equation \eqref{ZEqnNum528491}. The $\min \left(\cdot \right)$ finds the minimum distance among $\mathbf c_{j} $ to all $\mathbf c_{i} $, for \textit{i }= 1, 2, {\dots}, \textit{N}, \textit{i }$\mathrm{\neq}$ \textit{j}. The $\mathrm{mean}\left(\cdot\right)$ takes an average over all the minimum distances. The parameter \textit{$\beta$ }in the equation is for fine-tuning. 
\begin{equation} \label{ZEqnNum528491} 
	\hat{\sigma }=\beta {\mathop{\mathrm{mean}}\limits_{j}} \left({\mathop{\min }\limits_{i}} \left(\left\| \mathbf c_{i} -\mathbf c_{j} \right\| _{2} \right)\right).
\end{equation}

\subsection{Sequential amplitude estimation and demodulation}
We have not found an explicit expression of the secondary eigenvectors out of KPCA with respect to the steering vectors of the scatterers. Therefore, the proposed algorithm can only make use of the first column of $\mathbf{Y}$ that captures the steering vector of the most dominant scatterer well. The proposed algorithm employs a strategy that sequentially demodulates and estimates the most dominant contribution. However, a challenge arises, where the real variances of the scatterers are lost after the nonlinear transformation in KPCA, i.e., the real intensity of the scatterer is not represented by the eigenvalues of \textbf{Y}. Therefore, we estimate the intensity by the Rayleigh quotient \cite{Stoica1997Introduction}, i.e.,
\begin{equation} \label{ZEqnNum572230} 
	\hat{\sigma }_{1}^{2} =\frac{1}{N} \frac{\bar{\mathbf y}_{1}^{H} \hat{\mathbf C}\bar{\mathbf y}_{1} }{\bar{\mathbf y}_{1}^{H} \bar{\mathbf y}_{1} } ,  
\end{equation} 
where $\bar{\mathbf y}_{1} $ denotes the first column of $\mathbf{Y}$ with its amplitude dropped. Please note that the amplitude-dropped vector appears in the equation, instead of the normalized one that usually appears in the literature. Once the intensity of the scatterers is obtained, the covariance matrix can be demodulated as follows:
\begin{equation} \label{20)} 
	\hat{\mathbf C}_{update} =\hat{\mathbf C}-\hat{\sigma }_{1}^{2} \bar{\mathbf y}_{1} \bar{\mathbf y}_{1}^{H} .  
\end{equation} 
This KPCA plus demodulation process is iteratively performed until no more significant scatterers is left or the predefined maximum number of scatterers is reached. 

\subsection{Algorithm summary}

\begin{table}
	
	\caption{Summary of the proposed algorithm}
	\begin{tabular}{p{0.1cm}p{0.1cm}p{7.5cm}} 
		\hline 
		\multicolumn{3}{p{7.7cm}}{\textit{for} each pixel in the image} \\

		~&\multicolumn{2}{p{7.5cm}}{\textbf{Sample selection} (samples with similar layover configuration)} \\

		~&\multicolumn{2}{p{7.5cm}}{\textbf{Estimate sample coherence matrix} $\hat{\mathbf C}_{0} =\frac{1}{M} \mathbf{GG}^{H} $}\\
		
		~&\multicolumn{2}{p{7.5cm}}{\textbf{Center} $\hat{\mathbf C}_{0} $ yielding $\underline{\hat{\mathbf C}}_{0} $}\\
		
		~&\multicolumn{2}{p{7.5cm}}{\textbf{Estimate number of scatterer} \textit{K}, e.g., using minimum description length} \\
		
		~&\multicolumn{2}{p{7.5cm}}{\textit{for k = }scatterer 1 to \textit{K}}\\
		
		~&~&\textbf{Perform KPCA} on $\underline{\hat{\mathbf C}}_{k-1} $ according to equation \eqref{ZEqnNum873526}, obtaining dimension reduced data \textbf{Y} \\
		
		~&~&\textbf{Obtain estimate of steering vector }$\hat{\mathbf r}_{k} $\textbf{ }by taking the first eigenvector (corresponding to the largest eigenvalue) of \textbf{Y } \\
		
		~&~&\textbf{Estimate scatterer amplitude} $\sigma _{k} $ using Rayleigh quotient (equation \eqref{ZEqnNum572230}) \\
		
		~&~&(optional) \textbf{Re-estimate} $\sigma _{1} $ and $\sigma _{2} $ alternatingly when \textit{K }is only equal to two \\
		
		~&~&\textbf{Deflate} $\hat{\mathbf C}_{k-1} $, i.e., $\hat{\mathbf C}_{k} =\hat{\mathbf C}_{k-1} -\hat{\sigma }_{k}^{2} \hat{\mathbf r}_{k} \hat{\mathbf r}_{k}^{H} $ \\
		
		~&~&\textbf{Center} $\hat{\mathbf C}_{k} $ yielding $\underline{\hat{\mathbf C}}_{k} $ \\
		
		~&\multicolumn{2}{p{7.5cm}}{\textit{end}} \\
		
		~&\multicolumn{2}{p{7.5cm}}{\textbf{Height estimation} using standard PSI technique (e.g., periodogram or integer least square) from the estimated steering vectors $\left[\hat{\mathbf r}_{1} ,\hat{\mathbf r}_{2} ,\cdots \hat{\mathbf r}_{K} \right]$}\\
		
		\multicolumn{3}{p{7.7cm}}{\textit{end}} \\ 
		\hline 
		\label{tab:workflow}
	\end{tabular}
\end{table}

{The previous subsection describes the core of the proposed blind TomoSAR algorithm. However, }in TomoSAR applications, the processing pipeline shall also perform sample selection, model order selection, and height estimation, {besides the main steps of the proposed algorithm.} We summarize the full algorithm for a real SAR image processing in Table \ref{tab:workflow}. {Since the scope of this paper is mainly on mitigate the phase bias in the retrieve the steering vectors, we will not go into details of the other steps. The readers can refer to \cite{Zhu2015Joint, Deledalle2015NL-SAR} for sample selection, \cite{wang_robust_2016} for robust covariance matrix estimation, \cite{visuri_nonparametric_2000} for model order selection, and \cite{ferretti_permanent_2001} for final parameters estimation. }

\subsection{{Extension to D-TomoSAR}}
{Real data InSAR stacks are often multi-temporal, meaning the deformation shall also be considered. In the case of differential TomoSAR (D-TomoSAR), the reflectivity profile will be $\gamma \left(s\right)\delta\left(p_1-p_1\left(s\right),\cdots,p_M-p_M\left(s\right)\right)$ \cite{zhu_lets_2011}, where $p_i$ is the deformation axis and $p_i\left(s\right)$ is the deformation function along the elevation. The phase of the corresponding steering vectors is the sum of the height frequency as well as the deformation frequency. In discrete form, the dimension of $\mathbf{R}$ and $\bm\upgamma $ will increase according to the number of the deformation parameters, i.e., $\mathbf{R}\in {\mathbb C}^{N\times \left(L\times P_{1} \times P_{2} \cdots \right)} $, and $\bm\upgamma \in {\mathbb C}^{L\times P_{1} \times P_{2} \cdots } $, where \textit{P${}_{i}$} is the discretization level of the \textit{i}th deformation parameter.} 

{Since the height and deformation signal act on a scatterer at a fixed elevation location, the deformation and height signal of a scatterer can be considered as a single source, which can be retrieved using the proposed algorithm without any modification. The only necessary change in the whole processing pipeline is the last step (height estimation) in Table \ref{tab:workflow}. It will be a multi-dimensional periodogram, instead of one dimensional. The reader can refer to \cite{wang_robust_2018} for the application of the proposed algorithm on retrieving both the height and periodic deformation induced by thermal dilation.
}

\section{Experiments}

In this section, we evaluate the performance of the proposed method using simulated data, as well as real TerraSAR-X data.

\subsection{Simulation}

The performance of the proposed algorithm is firstly evaluated via simulations. As we are mainly interested in the phase of the scatterers, we define the angle between the estimated scatterer steering vector and the ground truth as the quality metric. The angle is defined as the arccosine of the inner product of the two vectors, which is shown as follows.
\begin{equation} \label{ZEqnNum210185} 
	b =\cos ^{-1} \left|\hat{\mathbf r}^{H} \mathbf r \right|,  
\end{equation} 
where both $\hat{\mathbf r} $ and $\mathbf r $ are assumed to be normalized. Because of the absolute value in equation \eqref{ZEqnNum210185}, the domain of the angular bias is [0$\mathrm{{}^\circ}$, 90$\mathrm{{}^\circ}$]. A wild guess of an unknown signal direction will be at 45$\mathrm{{}^\circ}$. Therefore, an angular bias greater than 45$\mathrm{{}^\circ}$ basically indicates a failed source separation.

The simulation setting was similar to that in Fig. \ref{fig:limit_pca_phase}. Since most of the TomoSAR literature assumes two dominant scatterers in TerraSAR-X data from urban areas, two-scatterer mixtures were simulated in our experiments. The number of images was set to 9. The baselines are equally distributed from -200m to 200m. The other parameters (e.g., wavelength, range distance, etc.) were set to be similar to those of TerraSAR-X. According to the analysis, the orthogonality and the variance of the signals heavily affects the performance of the BSS algorithms. Therefore, we vary the amplitude ratio, the distance between the two scatterers, and SNR in the simulation. {In the first two experiments, i.e. amplitude ratio and scatterer distance, we intend to test the phase bias of PCA and the proposed algorithm at noise free case. Therefore, we used 900 looks to estimate a very accurate covariance matrix. Of course, it is nearly identical by using the theoretical covariance matrix.} In all the experiments, we use 1000 times Monte Carlo simulation to estimate the mean and the standard deviation of the angular bias. 

\hfill

\noindent \textbf{Performance with respect to amplitude ratio}

\begin{figure}[h]
	\centering
	\includegraphics[width=0.48\textwidth]{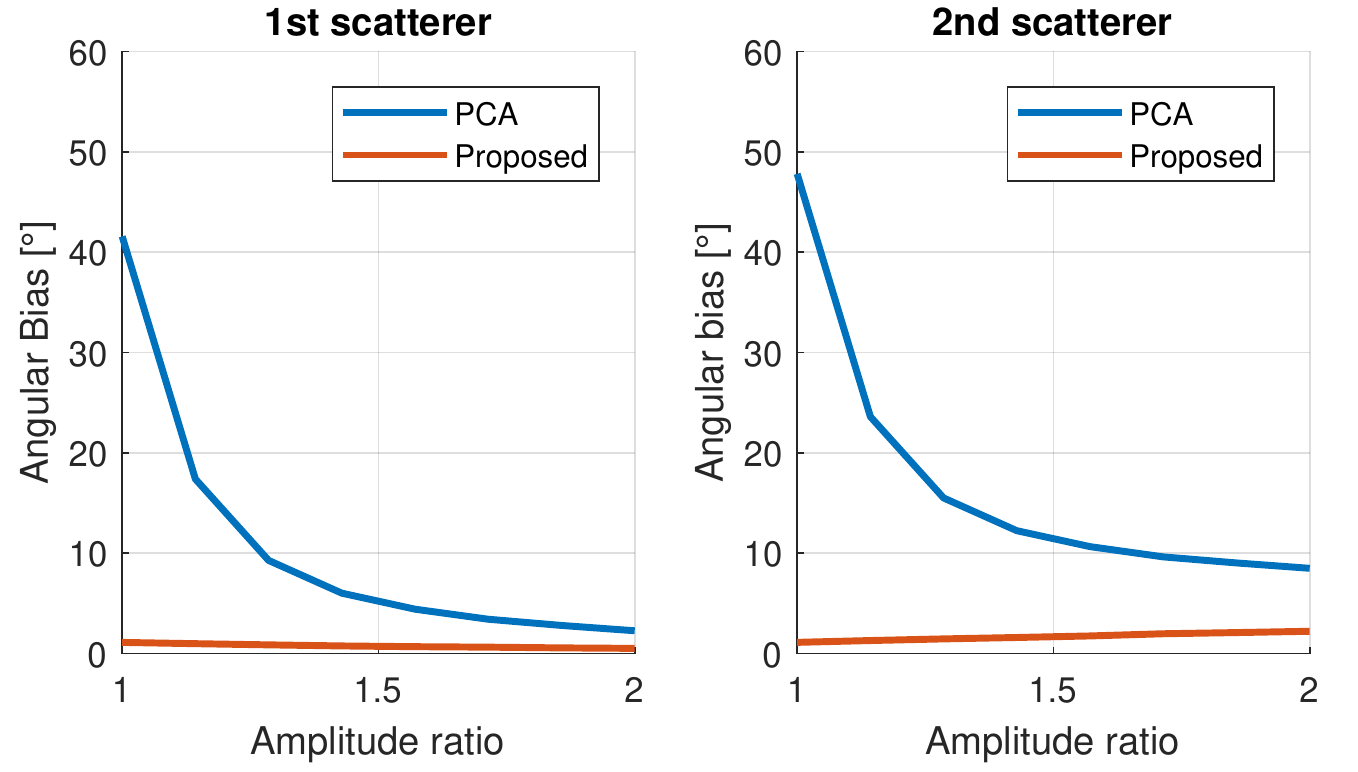}
	
	\caption{{The phase bias in degree of the estimated steering vectors of the first (brighter) scatterer (left), and the second scatterer (right) with respect to different amplitude ratio between the two scatterers. The elevation difference between the scatterers was set to one Rayleigh resolution (27.3m in the simulation). A Gaussian kernel was employed in the proposed algorithm. A total of 900 looks were used for covariance matrix estimation. No noise was introduced. The figure shows that PCA is sensitive to the intensity ratio of the two scatterers. A comparable intensity between the scatterer will almost result inseparation of the scatterers, even in a noise-free case.}}
	\label{fig:bias_amp_ratio}
\end{figure}

The first simulation set out to study the systematic bias of BSS algorithms and so no noise was included in the signal. The distance between the two scatterers was set to one Rayleigh resolution, which is 27.3 m in this case. Fig. \ref{fig:bias_amp_ratio} shows the angular bias of the steering vectors of the two scatterers estimated using PCA and the proposed method. The x-axis shows an increasing amplitude ratio from 1 to 2. As can be seen, very large phase bias appears in the PCA result when the amplitude ratio is low. As the amplitude ratio increases, the dominant direction becomes more prominent, and hence easier for PCA to capture. The same trend appears in the result of both the first and the second scatterers. In summary, PCA will not be a good choice for separating scatterers with similar brightness. However, the performance of PCA will be comparable to the proposed method when the amplitude ratio is larger than 2.

\hfill

\noindent \textbf{Performance with respect to scatterers distance }

\begin{figure*}[h]
	\centering
	\includegraphics[width=0.9\textwidth]{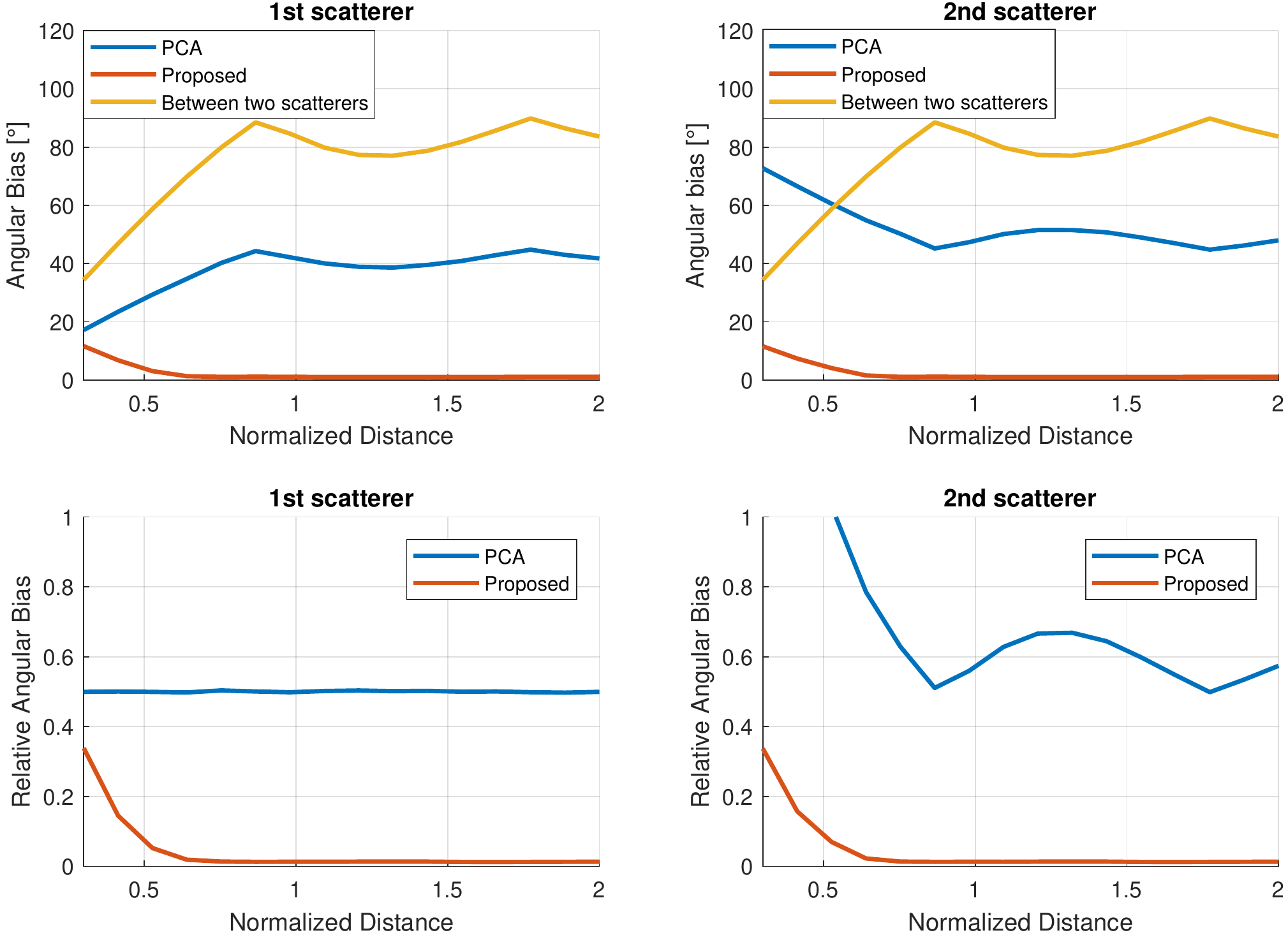}
	\caption{Upper row: The phase bias (in degree) of the estimated steering vectors of the first (brighter) scatterer (left), and the second scatterer (right) with respect to increasing distance between the two scatterers; lower row: the relative angular bias, which is the ratio of the angular bias and the angle between the true steering vectors of the two scatterers. The amplitude ratio of the scatterers was set to one. A Gaussian kernel was employed in the proposed algorithm. A total of 900 looks were used for covariance matrix estimation. No noise was introduced in the simulation.}
	\label{fig:bias_alpha}
\end{figure*}

In this experiment, we varied the distance between the scatterers from 0.3 to 2 Rayleigh resolution units. The amplitude ratio of the scatterers was kept at one. The upper row of Fig. \ref{fig:bias_alpha} shows the bias of the estimates with respect to increasing distance between the scatterers. For analysis, we also plotted the angle between the true steering vectors of the two scatterers in Fig. \ref{fig:bias_alpha}, which are shown as yellow curves. First, the proposed method greatly outperforms PCA in the whole range of scatterer distance. The angular bias of the first steering vector estimated by the proposed method stays at 2 to 3 degrees, whereas it is 40 degrees for PCA. Second, the performance of both methods stays relatively stable when the distance is larger than 0.8 Rayleigh resolution. Not surprisingly, the performance is clearly affected when it enters the super-resolution regime (i.e., distance shorter than one Rayleigh resolution). Interestingly, the results of PCA have a strong correlation with the original angle of the two scatterers. PCA, as a non-superresolving technique, will detect the first signal at a location between the two scatterers. Hence, the angular bias of the first signal decreases as the distance between the two scatterers decreases. This can be seen in the lower row of Fig. \ref{fig:bias_alpha}, which shows the ratio between the angular bias and the original angle between the two scatterers (steering vectors). In the left plot of the lower row of Fig. \ref{fig:bias_alpha}, the result of PCA stays at 0.5, indicating that PCA always detects the most dominant signal right in the middle of the two scatterers (when they are equally bright).

\hfill

\noindent \textbf{Performance with respect to SNR}

\begin{figure*}[h]
	\centering
	\includegraphics[width=0.85\textwidth]{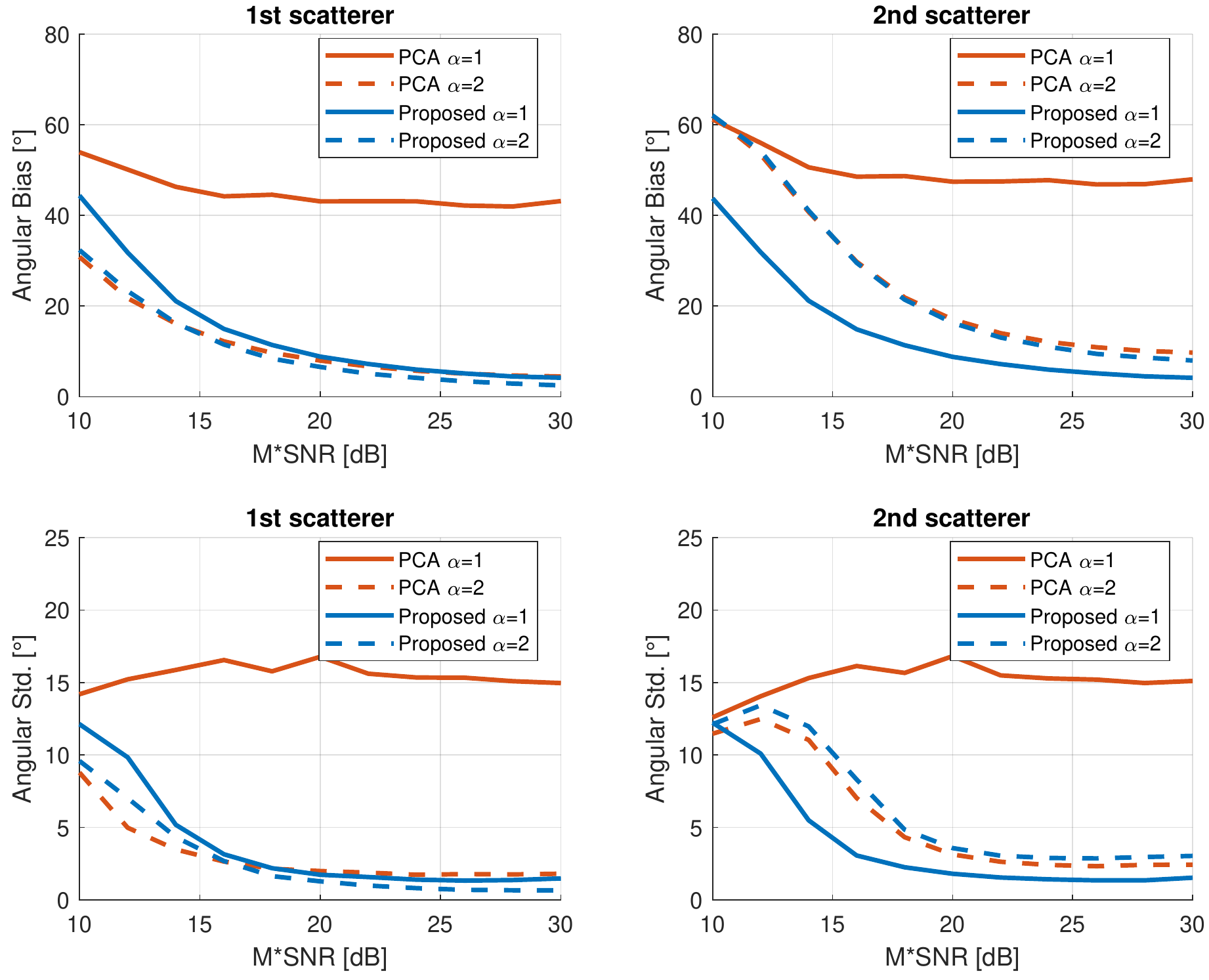}
	\caption{{The mean (upper row) and standard deviation (lower row) of the angular bias of the steering vector estimates with respect to $M\times SNR$. The solid curve corresponds to an amplitude ratio of 1, while the dashed curves correspond to a ratio of 2. The distance between the two scatterers was kept at one Rayleigh resolution. The result is consistent with Fig. \ref{fig:bias_amp_ratio} and Fig. \ref{fig:bias_alpha}. It is nearly impossible to separate two scatterers by PCA when the amplitude ratio is close to one, regardless of the SNR.}}
	\label{fig:bias_snr}
\end{figure*}

The previous experiments show that conventional PCA-based methods have strong systematic bias even under noise-free case. To evaluate the performance of the proposed algorithm under real scenario, we included additional complex circular Gaussian noise in the simulation. As the number of looks and SNR jointly affect the performance, their product was considered. Fig. \ref{fig:bias_snr} shows the mean and the standard deviation of the angular bias of the estimates with respect to to an increasing \textit{M}*\textit{SNR}. The solid curves shows the performance under a 1:1 amplitude ratio, whereas the dashed curves shows the result for the amplitude ratio of 2. First, the result is consistent with Fig. \ref{fig:bias_amp_ratio} and \ref{fig:bias_alpha}. It is nearly impossible to separate two scatterers using PCA when the amplitude ratio is close to one (red solid curve), regardless of the SNR. The performance of PCA is comparable to the proposed method when the amplitude ratio increases to 2. The SNR mildly affects the performance of the evaluated methods, especially when \textit{M}*\textit{SNR} is greater than 20dB. In modern high-resolution spaceborne SAR data, a 20 dB \textit{M}*\textit{SNR} is a rather relaxed condition, for example an SNR of 0dB with 100 looks.

\subsection{ Performance on real data}

\begin{figure*}[h]
	\centering
	\includegraphics[width=0.95\textwidth]{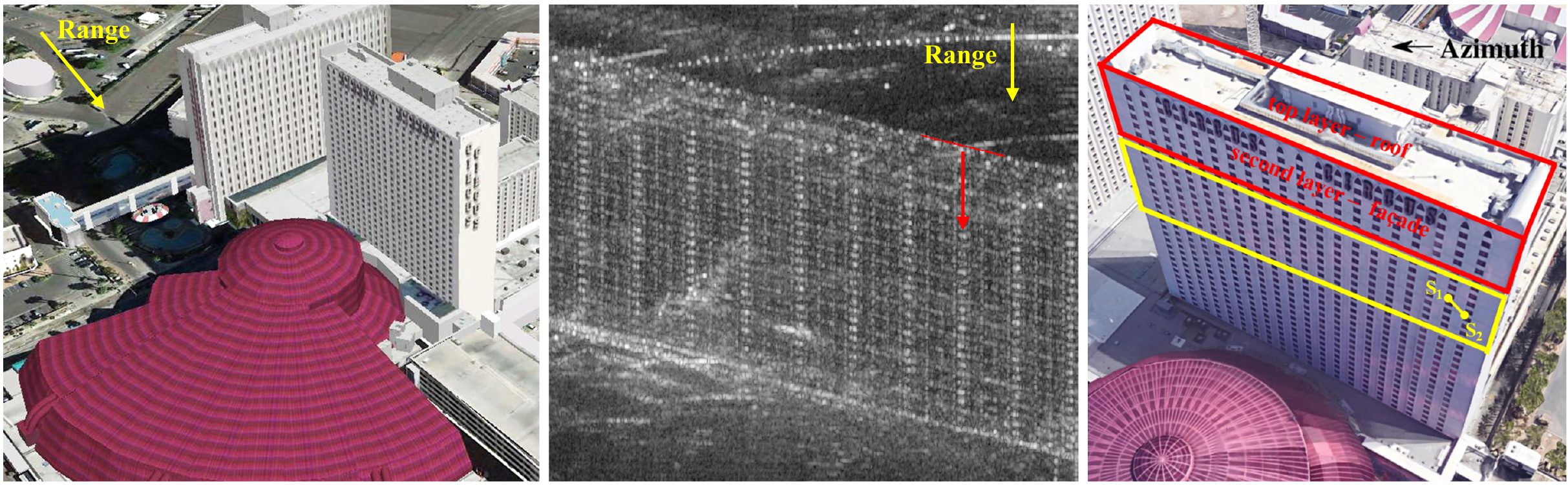}
	\caption{Left: Google image of the test building; middle: the mean amplitude image of the test building; and right: the layover configuration of the test building (Fig. cited from \cite{Zhu2015Joint}). 	The yellow arrows indicate the range direction. The red {line} on the amplitude image is a manually marked template of iso-height pixels. We assume that the pixels in the template have similar layover configuration, as well as similar scatterer statistics. 
	It slides down one pixel at a time. We estimate a single reflectivity profile from the covariance matrix of the pixels in the template at a given position. {In the experiment we also vary the length of the template to test the performance of the proposed algorithms under different number of looks.} 
	}
	\label{fig:real_data}
\end{figure*}

We tested our algorithm on six high-resolution TerraSAR-X interferograms acquired in pursuit-monostatic mode. The temporal baselines are in the order of seconds, so that the ground deformation and change in atmospheric phase are negligible. The optical and the SAR image of the test building are shown in Fig. \ref{fig:real_data}. The yellow arrows in the images indicate the range direction. {We manually define a iso-height direction parallel to the building floors, in order to guide the sample selection. This direction is shown as the thin red polygon} on the amplitude image. We assume that the pixels in the template have similar layover configuration, as well as similar scatterer statistics. This iso-height template slides down one pixel at a time. We estimate a single reflectivity profile from the covariance matrix of the pixels in this template at each given position. {In the experiment, we also vary the length of this template to test the performance of the proposed algorithm under different number of looks.}

Although this type of iso-height template was shown as an effective sample selection for joint height estimation in TomoSAR \cite{Zhu2015Joint}, the samples within the template cannot be guaranteed to be ergodic. Therefore, a robust covariance matrix estimator was employed in our processing. The sign covariance matrix (SCM) \cite{wang_robust_2016} is a non-iterative robust estimator, which is a good balance between robustness and computational cost. The SCM can be estimated as follows:
\begin{equation} \label{ZEqnNum978033} 
	\hat{\mathbf C}_{{\rm scm}} =\frac{1}{M} \sum _{m=1}^{M}\left\| \mathbf g_{m} \right\| _{2}^{-2} \mathbf g_{m} \mathbf g_{m}^{H}  ,  
\end{equation} 
where only the direction of each multivariate sample is considered. The real covariance is lost as a result.

The proposed algorithm was applied to the test building, with the template sliding from the top to the bottom of the building. For each template position, we extract the phase of the two most dominant components as the estimates of steering vectors. For conventional PCA, this will be the first two eigenvectors. For each estimate of steering vectors, a periodogram was calculated to estimate the corresponding elevation. In the experiments, we also tested the proposed algorithm at 50 looks and 500 looks, by varying the length of the template.

\begin{figure*}[h]
	\centering
	\includegraphics[width=0.95\textwidth]{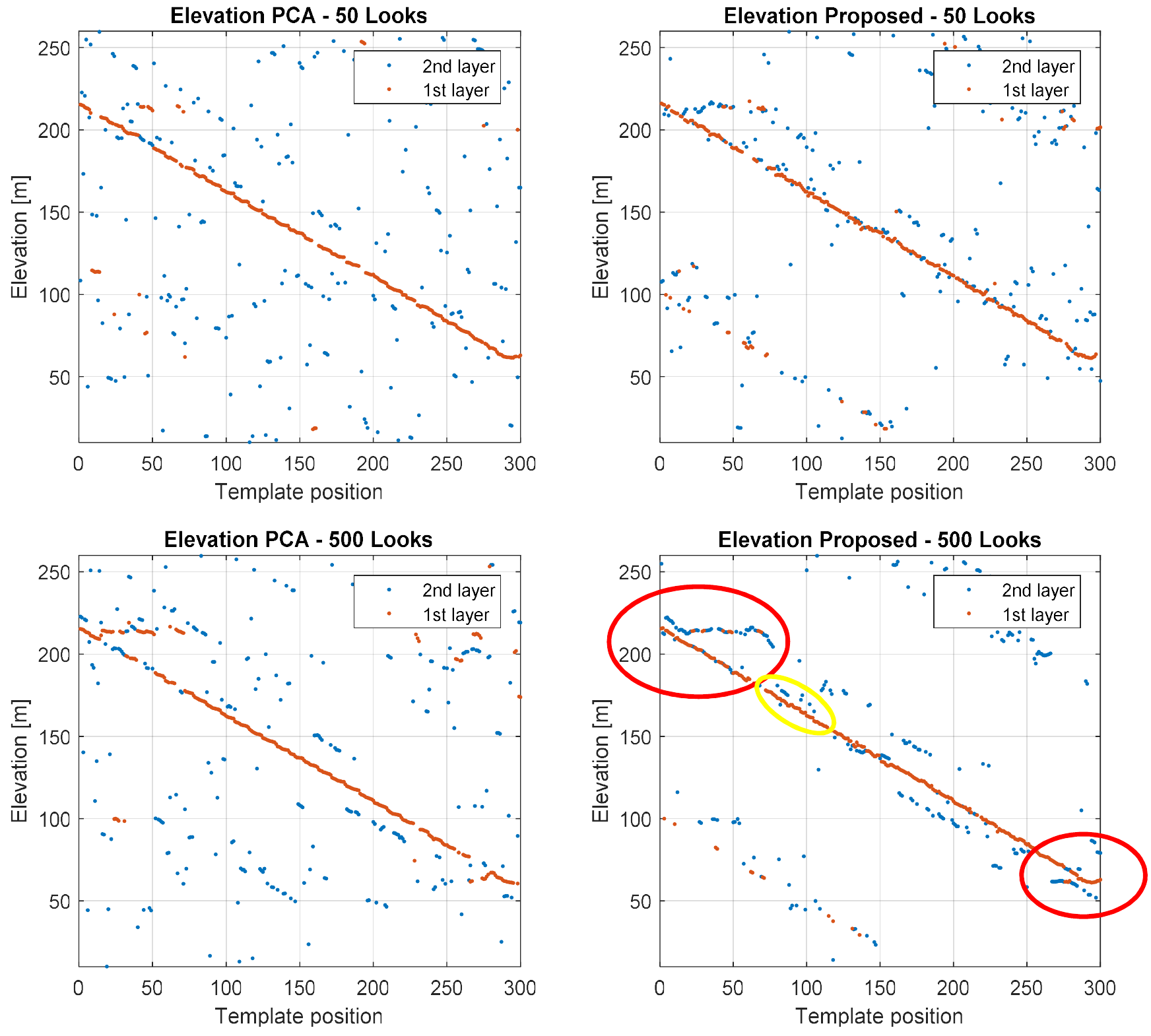}
	
	\caption{Elevation retrieved from the first two principle directions estimated by PCA (left), and by proposed algorithm (right). The red dots represent the first (brightest) layer, and the blue is the second layer. The red and yellow ellipses mark two layover regions on the test building. They correspond to the drawing in Fig. \ref{fig:real_data} (right).}
	\label{fig:result_real_data}
\end{figure*}

Figure \ref{fig:result_real_data} compares the two-layer elevation retrieved from PCA and the proposed algorithm. The red dots represent the first (brighter) layer, and the blue the second layer. {The upper row shows the results using only 50 looks, whereas the lower row is the result using 500 looks.} First, it is apparent that the proposed algorithm retrieves a much straighter fa\c{c}ade line, whereas small periodic undulations appears on the PCA result. This is likely due to the systematic bias caused by PCA. Second, fewer outliers appear in the results of the proposed method, showing a better robustness against sample nonergodicity. {However, the variance of fa\c{c}ade from the KPCA result shows slightly higher than that from PCA, indicating the proposed methods requires more number of looks than PCA.} By assuming the fa\c{c}ade to be a straight line, {the accuracy of PCA and the proposed algorithm are both ca. 1.0m using 50 looks. As the number of looks increases, the advantage of the proposed algorithm become more prominent, since the bias of the estimates becomes more prominent than the variance of the estimates. At 500 looks, the proposed algorithm outperforms PCA by a factor of 1.2 in the accuracy of height estimation.} Last but not least, the proposed algorithm was also able to retrieved very well the layover between the top fa\c{c}ade and the roof, as well as the layover between the lower fa\c{c}ade and the ground (marked by the red ellipse in Fig. \ref{fig:result_real_data}). The roof-fa\c{c}ade layover was also shown in the right of Fig. \ref{fig:real_data}.  We can see the minimum distance till which the two layers cannot be separated anymore is roughly 6m, which is below one Rayleigh resolution (11.6m in this data). 
These findings coincide with the finding in the simulation (Fig. \ref{fig:bias_alpha}) that shows the proposed method has extremely low phase bias, even in the super-resolution regime.

\section{ Discussion}

\subsection{ Kernel parameter selection}

The proposed algorithm requires the selection of a kernel parameter: either the polynomial order \textit{d} for a polynomial kernel, or the factor \textit{$\beta$} of the standard deviation in a Gaussian kernel. In order to obtain an operable range of those parameters, we measure the performance at different parameter settings by the ensemble coherence (periodogram) of the first vector of \textbf{Y} denoted by \textbf{y}${}_{1}$ with the true steering vector of the first scatterer denoted by \textbf{r}${}_{1}$. As only the phase is of interest to us, the ensemble coherence (equation \eqref{ZEqnNum134823}) is computed based on the amplitude-dropped vectors of \textbf{y}${}_{1}$ and \textbf{r}${}_{1}$${}_{,\ }$which are denoted by $\bar{\mathbf y}_{1} $ and $\bar{\mathbf r}_{1} $.
\begin{equation} \label{ZEqnNum134823} 
	\eta =\frac{1}{N} \left|\bar{\mathbf y}_{1}^{H} \bar{\mathbf r}_{1} \right| 
\end{equation} 
The ensemble coherence $\eta $ ranges from 0 to 1. A perfect reconstruction of the phase will result a coherence of one. 

\begin{figure}[h]
	\centering
	\includegraphics[width=0.48\textwidth]{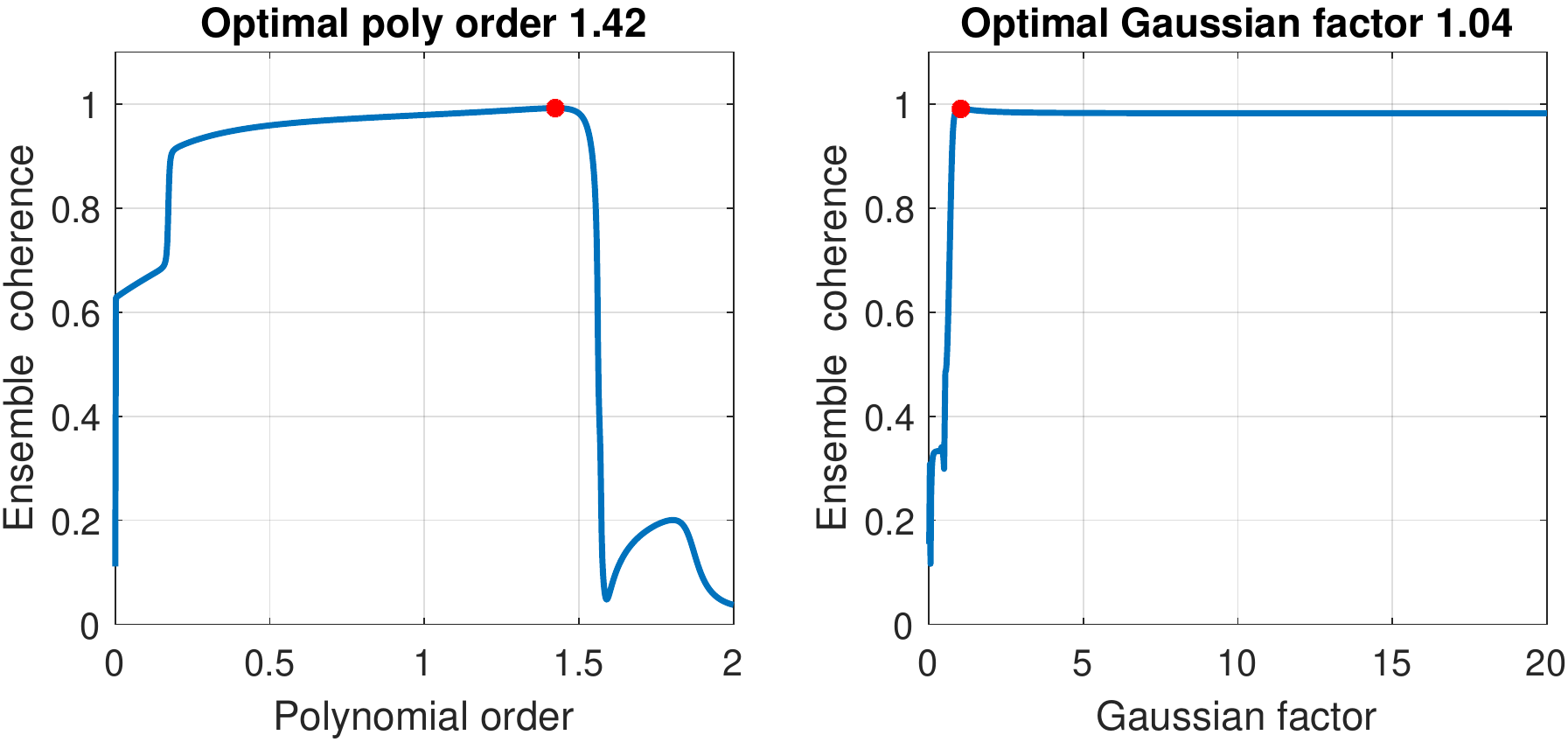}
	\caption{{The temporal coherence (complex correlation coefficient) between the first column of the dimension reduced data and the true steering vector of the brightest scatterer, with respect to different kernel parameter settings: order of polynomial kernel (left), and factor $\alpha$ (in equation \eqref{ZEqnNum528491}) of the standard deviation of the Gaussian kernel (right). A coherence of 1 refers to a perfect reconstruction. In the simulation, two scatterers are separated by one Rayleigh resolution, and have an amplitude ratio of 1.2.}}
	\label{fig:kernel_selection}
\end{figure}

Fig. \ref{fig:kernel_selection} (left) shows the ensemble coherence with respect to different polynomial orders applied to the simulated data used in Fig. \ref{fig:limit_pca_phase} (with $\alpha=1.2$). The experiment shows that the ensemble coherence stays at almost one for a wide range of polynomial orders, which indicates a wide operable range. The coherence drops rapidly beyond 1.5, which aligns with our hypothesis in Section \ref{sec:selection_kernel} that a higher polynomial order will introduce unwanted artificial scatterers with higher elevation. Our experiments show that a good starting point of the polynomial order is between 1.1 to 1.4. The operable range of $\beta$ can be examined using the same strategy. Experiment in Fig. \ref{fig:kernel_selection} (right) shows that the ensemble coherence stays at nearly one beyond certain values of $\beta$. This demonstrates that the operable range of $\beta$ is also very wide. In our experiments, we set $\beta=5$.

Optionally, those parameters can also be adaptively estimated. The polynomial order \textit{d} is mainly influenced by the elevation distance (angular difference) between the two scatterers because the polynomial order acts as a multiplication of the phase, whereas the parameter $\beta$ mainly depends on the amplitude ratio of the two scatterers as the ${L}_{2}$ norm in the Gaussian kernel is heavily governed by the signal magnitude. However, the elevation distance and amplitude ratio are both unknowns. One can start with an initial value of the kernel parameter, and refine it with the estimates of steering vectors and the amplitudes of the scatterers. In general, we found that a Gaussian kernel usually provides more stable performance than polynomial kernels in the experiments. The parameter $\beta$ is also less sensitive to the change in the data.

\subsection{Limitations of the proposed algorithm}

First, the proposed algorithm assumes a fully-coherent signal model. We have shown in Section \ref{sec:limit_pca_pca} that it does not work with the DS of an arbitrary coherence matrix. Additional constraints, such as coherence models, should be introduced in the proposed algorithm, in order to work with general DS models. However, our experiments do find the proposed algorithm works with a constant decorrelation matrix. 

We also found that the sample selection may be a challenge in real data processing. This selection is different from many statistical tests mentioned in SqueeSAR or similar articles \cite{Ferretti2011New,Parizzi2011Adaptive,Wang2012Retrieval}. Therefore, we manually drew an iso-height template in the experiment. For a fully automatic processing, we shall incorporate available GIS building footprints, or resort to methods like NL-SAR \cite{Deledalle2015NL-SAR}. However, these actions (especially NL-SAR) will increase the computational cost, which counteracts the motivation of the proposed algorithm. 

Last, the proposed algorithm shows certain super-resolution capability. However, we must note that the knowledge of two scatterers was given in the experiments, meaning no model selection/detection was performed. Therefore, integrating a full detection step to the proposed algorithm is required for a complete assessment of its super-resolution power.

\section{Conclusion}

This article proposed a robust method to blindly perform layover separation in multibaseline InSAR data. Such blind separation requires no inversion of the SAR imaging model, hence reducing the computation cost logarithmically. The proposed algorithm outperforms the state of the art in various aspects. We showed that the state of the art could obtain a reasonable result only under good orthogonality conditions, i.e., large elevation and amplitude difference between the scatterers. The proposed method employs KPCA to artificially increase the dimension of the data, hence achieving superior performance. Simulation shows that the proposed algorithm is nearly optimum in the common range of SNR and number of looks. Experiments on real data show that the proposed method outperforms the state-of-the-art method by a factor of one to three in terms of the height accuracy, depending on the used number of looks. Surprisingly, the proposed method is also capable of achieving a reasonable super-resolution capability, which is not shown in algorithms of the same kind.

This article shows a perspective on the data-driven approach of multibaseline InSAR algorithms. The long-term goal is to make good use of massive globally available SAR data. One immediate objective is to focus on an automatic and efficient sample selection strategies, such as incorporating freely available GIS building footprints. To further study data-driven approaches, we shall research subspace learning approaches in general, with application to TomoSAR. In addition, we can also investigate other emerging  data-driven alternatives, such as deep learning to reduce the computational cost of TomoSAR parameter optimization.


%




\ifCLASSOPTIONcaptionsoff
  \newpage
\fi



\bibliographystyle{IEEEtran}
\bibliography{ref}
%



%

\begin{IEEEbiography}[{\includegraphics[width=1in,height=1.25in,clip,keepaspectratio]{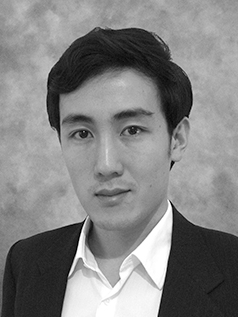}}]{Yuanyuan Wang}
received the B.Eng. degree (Hons.) in electrical engineering from The Hong Kong Polytechnic University, Hong Kong, in 2008, and the M.Sc. and Dr. Ing. degree from the Technical University of Munich (TUM), Munich, Germany, in 2010 and 2015, respectively. In June and July 2014, he was a Guest Scientist with the Institute of Visual Computing, ETH Zürich, Zürich, Switzerland. He is currently with the Department of EO Data Science, Remote Sensing Technology Institute, German Aerospace Center, Weßling, Germany, where he leads the working group Big SAR Data. His research interests include optimal and robust parameters estimation in multibaseline InSAR techniques, multisensor fusion algorithms of synthetic aperture radar (SAR) and optical data, nonlinear optimization with complex numbers, machine learning in SAR, and high-performance computing for big data. Dr. Wang was one of the best reviewers of the IEEE Transactions on Geoscience and Remote Sensing in 2016. He is a Member of the IEEE.
\end{IEEEbiography}

\begin{IEEEbiography}[{\includegraphics[width=1in,height=1.25in,clip,keepaspectratio]{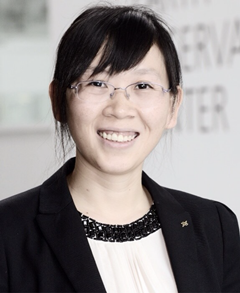}}]{Xiao Xiang Zhu}(S'10--M'12--SM'14) received the Master (M.Sc.) degree, her doctor of engineering (Dr.-Ing.) degree and her “Habilitation” in the field of signal processing from Technical University of Munich (TUM), Munich, Germany, in 2008, 2011 and 2013, respectively.
\par
She is currently the Professor for Signal Processing in Earth Observation (www.sipeo.bgu.tum.de) at Technical University of Munich (TUM) and the Head of the Department ``EO Data Science'' at the Remote Sensing Technology Institute, German Aerospace Center (DLR). Since 2019, Zhu is a co-coordinator of the Munich Data Science Research School (www.mu-ds.de). Since 2019 She also heads the Helmholtz Artificial Intelligence Cooperation Unit (HAICU) -- Research Field ``Aeronautics, Space and Transport". Since May 2020, she is the director of the international future AI lab "AI4EO -- Artificial Intelligence for Earth Observation: Reasoning, Uncertainties, Ethics and Beyond", Munich, Germany. Prof. Zhu was a guest scientist or visiting professor at the Italian National Research Council (CNR-IREA), Naples, Italy, Fudan University, Shanghai, China, the University  of Tokyo, Tokyo, Japan and University of California, Los Angeles, United States in 2009, 2014, 2015 and 2016, respectively. Her main research interests are remote sensing and Earth observation, signal processing, machine learning and data science, with a special application focus on global urban mapping.

Dr. Zhu is a member of young academy (Junge Akademie/Junges Kolleg) at the Berlin-Brandenburg Academy of Sciences and Humanities and the German National  Academy of Sciences Leopoldina and the Bavarian Academy of Sciences and Humanities. She is an associate Editor of IEEE Transactions on Geoscience and Remote Sensing.
\end{IEEEbiography}





\end{document}